\documentclass[twocolumn,aps]{revtex4}
\usepackage[english]{babel}
\usepackage{amssymb}
\usepackage{latexsym}
\usepackage{epsfig}
\usepackage{color}
\usepackage{float}
\usepackage{graphicx}
\usepackage{epstopdf}
\usepackage{natbib}
\usepackage{amsmath}
\usepackage[hidelinks=true]{hyperref}
\hypersetup{
  colorlinks   = true, 
  urlcolor     = red, 
  linkcolor    = blue, 
  citecolor   = red 
}

\begin{document}

\title{Study of Tsallis holographic dark energy model in the framework of Fractal cosmology}
\author{Abdulla Al Mamon\footnote{abdulla.physics@gmail.com}}
\affiliation{Department of Physics, Vivekananda Satavarshiki Mahavidyalaya (affiliated
to the Vidyasagar University), Manikpara-721513, West Bengal, India}

\begin{abstract}
In this work,  we study the evolution of a fractal universe composed of Tsallis holographic dark energy (THDE) and a pressureless dark matter that interact with each other through a mutual interaction. We then reconstruct the interaction term of this model by considering the Hubble length as the IR cut-off scale. We also study the behavior of different cosmological parameters during the cosmic evolution from the early matter-dominated era until the late-time acceleration. The present study shows that the universe undergoes a smooth transition from a decelerated to an accelerated phase of expansion in the recent past. Moreover, we also shown the evolution of the normalized Hubble parameter for our model and compared that with the latest cosmic chronometer data. Finally, we test the viability of the model by exploring its stability against small perturbation by using the squared
of the sound speed.
\end{abstract}

\maketitle

\section{Introduction}
Many observational datasets \cite{acc1,acc2,acc3,acc4,acc5,acc6,acc7} clearly indicate a mysterious type of energy with negative pressure, namely dark energy (DE), is needed to describe the late-time accelerated expansion phase of the universe. Different kinds of theoretical models have already been constructed to interpret accelerating universe and some eminent reviews on this topic can be found in \cite{de1,de2,de3,de4,de5}.
However, the problem of the onset and nature of this acceleration mechanism remains an open challenge of modern cosmology. \\

\par Recently, it has been proposed that the generalized entropy formalism can be used to study the cosmological and gravitational and  phenomena~\cite{non2,non4,non5,non6,non7,CI,5,non13,non19,non20,
non21,GSM,GSM1,GSM2,GSM3,smm,epjcr,Tsallis}. In this context, an alternative DE model has been proposed by using the Tsallis entropy \cite{Tsallis} and the holographic hypothesis \cite{hooft,sussk,HDE,li}, named Tsallis holographic dark energy (THDE) \cite{Tavayef}. The cosmological features of this DE model in different cosmological setups can be found in  \cite{Tavayef,thde1,thde2,thde3,thde4,thde5,thde6,thde7,thde8,thde9}. The energy density of THDE is given by \cite{Tavayef}
\begin{eqnarray}\label{THDED}
\rho_{D}=\frac{3}{8\pi}BL^{2\delta-4},
\end{eqnarray} 
in which $B$ is an unknown parameter, $L$ is the IR cutoff and $\delta$ is a free parameter. It is worthy to mention that for $\delta=1$. the above THDE reduces to the
holographic DE model.\\ 

\par In this context, it deserves to mention here that observations also allow a mutual interaction between the dark sectors (dark energy and dark matter (DM)) of the cosmos can solve the cosmic coincidence problem~\cite{Int0,Int00,Int1,Int10,Int11} (for review, see \cite{Bolotin} and references therein). Recently, holographic dark energy models, in the framework of fractal universe, have gained interest to explain the late-time cosmic acceleration \cite{thde9,Karami,Salti,Saavedra,Sadri,Bolotin}.\\

\par Following \cite{thde9}, in the present work, we are also
interested in studying the dynamics of a fractal universe filled with a THDE and pressureless matter  in an interacting scenario. In particular, we study the evolution of different cosmological parameters by considering an interaction between DM
and THDE whose IR cutoff is the Hubble horizon. However, the present work is 
different from the work \cite{thde9} in different ways. Here, we study
consequences of the interacting model in more details. We also study the evolution of the normalized Hubble parameter for our model and the standard $\Lambda$CDM model and compare that with the observational Hubble parameter data obtained through the cosmic chronometer method. Finally, we also study the stability of the model against small perturbations by using the squared of the sound speed.\\ 
\par The manuscript is organized as follows. In the next section, we describe the theoretical framework of our study. Here, we also present some general features of the proposed model. In section \ref{sec-result} we discuss the results of the model. Finally, section \ref{sec-conclu} is devoted to conclusions. \\
\par Throughout the text, we use units such that $G=c=\hbar=1$. 
\section{The interacting THDE model in the framework of Fractal Cosmology}
The action of Einstein gravity in fractal space-time is given by \citep{Frac1}
\begin{eqnarray}\label{action}
{\mathcal S}=\int d^4x\sqrt{-g}\Big(\frac{R-\omega\partial_\mu\nu\partial^\mu\nu}{16\pi }+{\mathcal L}_m\Big),
\end{eqnarray}
where, $g$, $R$, $\omega$, $\nu$ and $ {\mathcal L}_m $ are the determinant of the metric tensor $ g_{\mu\nu} $, Ricci curvature scalar, fractal parameter, fractal function and matter part of total Lagrangian density, respectively. The first Friedmann equation in a flat fractal universe corresponding to the action~(\ref{action}) is then obtained as \citep{Frac1}
\begin{eqnarray}\label{Fried1}
H^2+H\frac{\dot{\nu}}{\nu}-\omega\frac{\dot{\nu}^2}{6}=\frac{1}{3M_p^2}(\rho_m+\rho_D),
\end{eqnarray}
in which the fractal function is chosen in a power-law form as, $ \nu=a^{-\beta} $ with $\beta>0$ \citep{Frac1,Frac2,Frac3} and $ M_p^{-2}$ is the reduced Planck mass. In the above equation, $ H=\frac{\dot{a}}{a} $ is the Hubble parameter and an overdot denotes derivative with respect to cosmic time. Also, $\rho_m $ and $ \rho_D $ represent the energy densities of DM and THDE, respectively.\\
\par Here, we also assume that there is mutual interaction between the dark sectors (i.e., DM and THDE) of the fractal universe, then the conservation equations for DM and THDE
are given by
\begin{eqnarray}\label{DMCons}
\dot{\rho}_m+(3-\beta)H\rho_m=Q,
\end{eqnarray}
\begin{eqnarray}\label{DECons}
\dot{\rho}_D+(3-\beta)(1+\omega_D)H\rho_D=-Q,
\end{eqnarray}
where, $w_D=\frac{p_D}{\rho_D}$ and $p_D $ represent the equation of state (EoS) parameter and the pressure of the THDE, respectively. Also, the quantity $Q$ represents the interaction between these dark sectors. In fact, there are many proposed interactions in the literature to study the dynamics of the universe and for review, one can look into \cite{Bolotin} and references therein. However, in a recent work, the authors studied various phenomenological linear and non-linear interaction cases
in the framework of the holographic Ricci DE model and their investigation shows that the linear interaction $Q =3Hb^{2}\rho_D$ is the best case among the others (for details, see \citep{intcomp}). Motivated by these facts, in the present work, we assume 
\begin{equation}\label{anzQ}
Q =3Hb^2\rho_D
\end{equation}
in which $ b^2 $ is a coupling constant. It is clear from the conservation equations (\ref{DMCons}) and (\ref{DECons}) that $Q > 0$ indicates an energy transfer from the THDE to the DM, while for $Q < 0$, the energy transfers from the DM to the THDE. On the other hand, if $Q = 0$, then the THDE and DM evolve separately. So, the
sign of $Q$ will determine the direction of energy flow between the dark sectors. In this context, the interacting term $Q$, as given in equation (\ref{anzQ}), deserves further investigation.\\

\par The Friedmann equation (\ref{Fried1}) can
be written, in terms of density parameters, as 
\begin{eqnarray}\label{Fried2}
\Omega_m+\Omega_D=\gamma +1,
\end{eqnarray}
with  
\begin{equation}
\Omega_m=\frac{8\pi}{3H^2}\rho_m,
\end{equation}
\begin{equation}\label{Omega}
\Omega_D=\frac{8\pi}{3H^2}\rho_D,
\end{equation}
\begin{equation}
\gamma=-\beta-\frac{\beta^2\omega}{6}a^{-2\beta},
\end{equation}
\noindent Here, we consider the
Hubble horizon $H^{-1}$ as the IR cutoff $L$, then the energy
density of THDE (\ref{THDED}) is obtained as
\begin{equation}\label{rho1}
\rho_D=\frac{3}{8\pi}BH^{4-2\delta},
\end{equation}
Now, using the above equation, we re-expressed equation (\ref{Omega}) as
\begin{equation}\label{eqnh}
h=\frac{H}{H_0}=\left(\frac{\Omega_{D}}{\Omega^{0}_{D}}\right)^{\frac{1}{2(1-\delta)}},
\end{equation}
where, $h$ is the normalized Hubble parameter, $\Omega_{D0}$ is the current value of $\Omega_D$ and $H_{0}=\left(\frac{\Omega_{D0}}{B}\right)^{\frac{1}{2(1-\delta)}}$, represents the current value of $H$.\\
\par Now, taking the time derivative of equation (\ref{Fried1}),
along with using equations (\ref{DMCons}), (\ref{Fried2}) and (\ref{rho1}), we obtain 
\begin{equation}\label{dotH}
\frac{\dot{H}}{H^2}=\\
\frac{(3b^2-\beta+3)\Omega_D+(\beta-3)(1+\gamma)-\frac{\beta^3\omega}{3}(1+z)^{2\beta}}
{(2\delta-4)\Omega_D+2(1-\beta)-\frac{\beta^2\omega}{3}(1+z)^{2\beta}},
\end{equation}
where, $z=\frac{1}{a}-1$, is the redshift parameter.

Similarly, taking the time derivative of Eq.~(\ref{rho1}) and combining the result with
equations (\ref{DECons}) and (\ref{Fried2}), one can easily obtain
\begin{equation}\label{EoS11}
\omega_D=-1-\frac{3b^2}{3-\beta}+\frac{2\delta-4}{3-\beta}\frac{\dot{H}}{H^2},
\end{equation}
The equation of motion for the dimensionless THDE density parameter $\Omega_D$ can be obtained by differentiating equation (\ref{Omega}) with respect to the cosmic time and using equation (\ref{EoS11}). The result is
\begin{eqnarray}\label{dotOmega1}
&&{\Omega}^\prime_D=\frac{d\Omega_D}{d{\rm ln}a}=\Omega_D(1-\delta)\times\\&&
\frac{\Big((3b^2-\beta+3)\Omega_D+(\beta-3)(1+\gamma)-\frac{\beta^3\omega}{3}(1+z)^{2\beta}\Big)}
{(\delta-2)\Omega_D-\frac{\beta^2\omega}{6}(1+z)^{2\beta}-\beta+1},\nonumber
\end{eqnarray}
The deceleration parameter is defined as
\begin{equation}
q=-\frac{\ddot{a}}{aH^2}=-1-\frac{\dot{H}}{H^2},
\end{equation}
and the expression for $\frac{\dot{H}}{H^2}$ is given in equation (\ref{dotH}). Note that the expressions of $\omega_D$, $\Omega_D$ and $q$ are similar to the
results of \cite{Tavayef} for $ \beta=0 $ and $ b=0 $. Now, the total EoS parameter is given by
\begin{equation}
\omega_{tot}=-1-\frac{2\dot{H}}{3H^2}=-\frac{1}{3}+\frac{2q}{3},
\end{equation}
As is well known, $\omega_{tot}<-\frac{1}{3}$ is require to accelerate the
expansion of our universe. \\
Finally, we also check the viability of the THDE model by exploring its stability against small perturbation. In this context, we derive the squared speed of sound defined by
\begin{equation}
v^{2}=\frac{{\dot{p}}_D}{{\dot{\rho}}_D}=\omega_{D}+\frac{{\rho}_D}{{\dot{\rho}}_D}{\dot{\omega}}_D
\end{equation}
The model is classically stable if $0<v^{2}<1$.
\section{Results and discussion}\label{sec-result}
\begin{figure}[!]
\begin{center}
    \includegraphics[width=7cm]{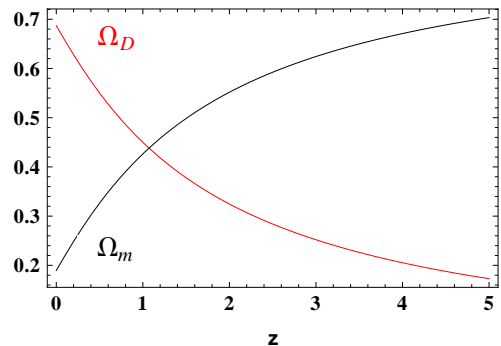}
    \caption{Plots of $\Omega_D$ (red curve) and $ \Omega_m$ (black curve) as a function of $z$ are shown for the best-fit values of model parameters ($\Omega_{D0},\delta,\omega,b^{2},\beta$), as given in table \ref{table1}.}\label{figdenp}
\end{center}
\end{figure}

\begin{figure}[!]
\begin{center}
    \includegraphics[width=7cm]{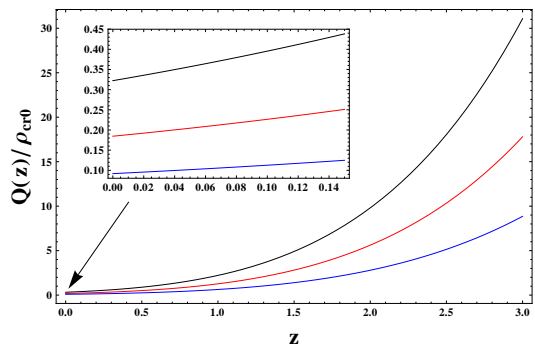}
    \caption{The plot of $Q$ (in units of critical density $\rho_{cr0}$) versus $z$ is shown for the present model (\ref{anzQ}) by considering the best-fit values of model parameters given in table \ref{table1} and different values of B. The black, red and blue curves are for $B=$ 1.1, 0.9 and 0.7, respectively.}\label{figint}
\end{center}
\end{figure}

The evolutions of the density parameters $\Omega_m$ and $ \Omega_D$ against the redshift parameter $ z $, according to the best fitted values of the parameters given in table \ref{table1}, is plotted in figure~\ref{figdenp}. From this figure, we observe that at high redshift, $\Omega_m$ dominates over $\Omega_D$, while at the late time, $\Omega_D$ dominates over $\Omega_m$. On the other hand, figure \ref{figint} shows the nature of the interaction term $Q$ given in equation (\ref{anzQ}). It has been found that $Q$ is positive through the evolution and thus the energy transfers from THDE to DM which
is well consistent with the Le Chatelier-Braun principle and the second law of thermodynamics (for instance, see\cite{lcbpavon}). It is also evident from figure \ref{figint} that the amount of energy flow is very less at recent time, but it was significantly high at earlier time. This is the reason that in spite
of the fact that energy flows from the THDE to the DM component, the evolution dynamics of the universe for the present model is such that $\Omega_D$ dominates over $\Omega_m$ at the current epoch (figure \ref{figdenp}). This feature provides a possible way out for the coincidence problem.\\
\par
\begin{figure}[!]
\begin{center}
    \includegraphics[width=7cm]{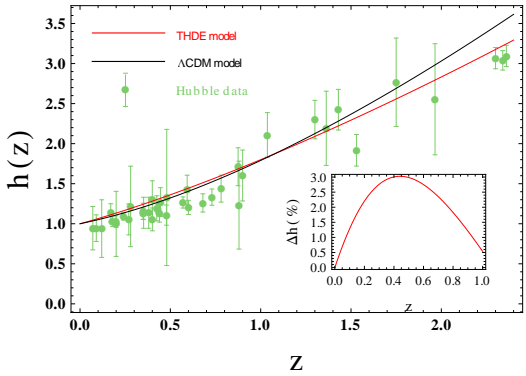}
    \caption{Comparison of the observed $H(z)$ data consisting 41 data
points \cite{nh1,nh2} and theoretical evolution of the normalized Hubble parameter, $h(z)=\frac{H(z)}{H_0}$, using the best-fit values of free parameters, as given in table \ref{table1}. Also, the corresponding error in $h(z)$ 
is given as \cite{aamsighma}, $\sigma_{h}=h\sqrt{\frac{\sigma^{2}_{H_0}}{H^{2}_0} + \frac{\sigma^{2}_{H}}{H^{2}}}$, with $\sigma^{2}_{H}$, $\sigma^{2}_{H_0}$ are errors
in $H$ and $H_0$ respectively. For the standard $\Lambda$CDM model, we taken $\Omega_{m0}=0.315$ from \cite{h4}. Also, the relative
deviation $\bigtriangleup h(\%)=\frac{h(z)-h_{\Lambda CDM}(z)}{h_{\Lambda CDM}(z)}\times 100$, is shown in the inner panel of the figure.}\label{fighzdh}
\end{center}
\end{figure}

 Figure \ref{fighzdh} shows the evolution of the normalized Hubble parameter $h$ as a function of $z$ for the THDE model and the flat $\Lambda$CDM, and compared them with the data points which have been obtained from the latest compilation of 41 data points of $H(z)$ measurements \cite{nh1,nh2}. One can easily observe from figure \ref{fighzdh} that the interacting THDE model reproduces the observed values of $h(z)$ quite well for each data point.\\ 
\par
\begin{figure}[!]
\begin{center}
    \includegraphics[width=7cm]{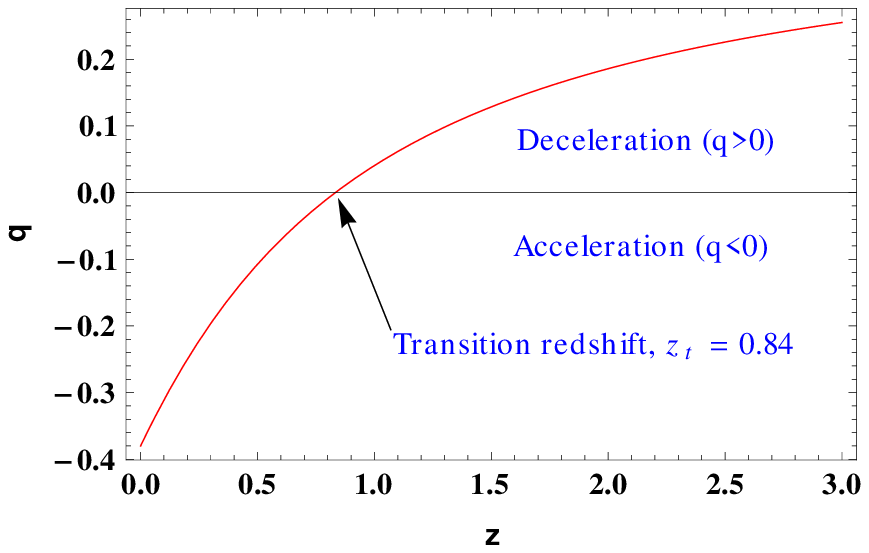}\\
    \vspace{5mm}
    \includegraphics[width=7cm]{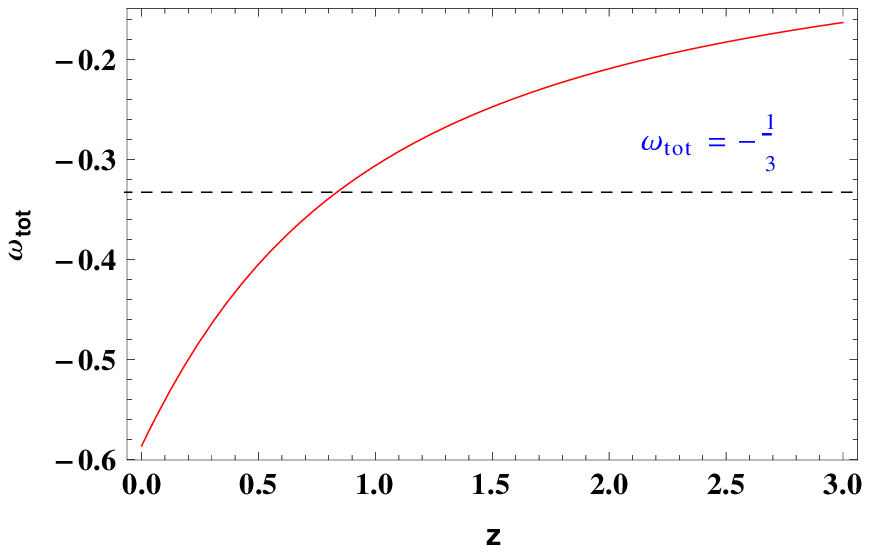}
    \caption{Upper panel: The evolution of the deceleration parameter, as a function of the redshift $z$, for the same values of the free parameters as given in figure \ref{figdenp}. The horizontal line stands for $q(z) = 0$. Lower panel: The evolution of the corresponding
total equation of state parameter $\omega_{tot}$. Note that the horizontal dashed line stands for $\omega_{tot}=-\frac{1}{3}$.}\label{figq}
\end{center}
\end{figure}

\begin{figure}[!]
\begin{center}
    \includegraphics[width=7cm]{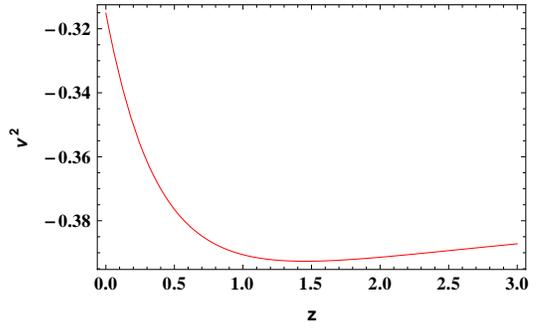}
    \caption{The evolution of the squared sound speed $v^{2}$, as a function of the redshift $z$, for the same values of the model parameters as given in figure \ref{figdenp}.}\label{figv2}
\end{center}
\end{figure}

 For a completeness, we have also reconstructed the evolutions of the deceleration parameter $q(z)$ and the total equation of state parameter $\omega_{tot}(z)$
for the present model. The plot of $q$ versus redshift $z$ is shown in the upper panel of figure \ref{figq}, while the corresponding plot of $\omega_{tot}(z)$ is shown in the upper panel of figure \ref{figq}. We observed that the THDE model exhibits a smooth transition from early deceleration era to the present acceleration era of the universe at the transition redshift $z_{t}=0.83$ for best-fit values of model parameters. This is in accordance with the current cosmological observations ($0.5<z_{t}<1$) \cite{zt1,zt2,zt3,zt4,zt5,zt6}. We also found from the
lower panel of figure \ref{figq} that $\omega_{tot}(z)$ was very close to
zero at high redshift and attains negative value $(-1 < \omega_{tot} <-\frac{1}{3}$)
at low redshift and further remains always greater than $-1$. Consequently, this does not suffer from the problem of `future singularity'. Finally, in order to test the classical stability of the THDE model, we also plot the square of sound speed in figures \ref{figv2}. From this figure, it has been observed that the model is unstable ($v^{2}<0$).
\begin{table}[H]
\centering
\caption{Best fit values of the model parameters with $1\sigma$ error bars obtained in
\cite{thde9} by using the combined (Pantheon SNIa+BAO+CMB+GRB) dataset, for the present model.}
\label{table1}
\hspace*{-1.5em}
\footnotesize\addtolength{\tabcolsep}{3pt}
\begin{tabular}{|c|c|}
\hline\hline
\multicolumn{2}{c}{ }                                             \\ [-0.4cm]
\multicolumn{1}{c}{Parameters}                 & \multicolumn{1}{c}{}           \\ [-0.02cm]\hline
\multicolumn{1}{c}{}                              & \multicolumn{1}{l}{}                          \\  [-0.2cm]
\multicolumn{1}{c}{$H_0$ }   & \multicolumn{1}{c}{$68.783^{+0.961}_{-0.761}$}   \\ [0.15cm]
\multicolumn{1}{c}{$\Omega_{D0}$ }             & \multicolumn{1}{c}{$0.687^{+0.024}_{-0.028}$} \\ [0.15cm]
\multicolumn{1}{c}{$\delta$}      & \multicolumn{1}{c}{$1.360^{+0.160}_{-0.191}$}     \\ [0.15cm]
\multicolumn{1}{c}{$\omega$}      & \multicolumn{1}{c}{$0.201 _ { - 0.029 } ^ { + 0.029 }$} \\[0.15cm]
\multicolumn{1}{c}{$b^2$}      &  \multicolumn{1}{c}{$0.0423 _ { - 0.02 } ^ { + 0.02}$} \\ 
 [0.15cm]
\multicolumn{1}{c}{$\beta$}      & \multicolumn{1}{c}{$0.123^{+0.059}_{-0.063}$}  \\ 
 \hline\hline
\end{tabular}
\end{table}
\section{Conclusions}\label{sec-conclu}
Here, we studied an interacting Tsallis holographic dark energy (THDE) model with Hubble horizon as IR cutoff in the framework of the flat fractal universe. Under this scenario, we then drived the deceleration parameter, the density parameter and the total equation of state parameter for the THDE. The resulting cosmological
scenarios are found to be very interesting. For obtaining
the results, we have considered the best-fit values of the free parameters ($H_{0}, \Omega_{D0}, \delta, \omega, b^{2}$ and $\beta$) obtained in \cite{thde9} by using the combined (Pantheon SNIa+BAO+CMB+GRB) dataset. The main findings of our model are given as follows.\\

\par We found that the deceleration parameter demonstrates a universe with
accelerating rate of expansion and it can be seen that the THDE model enters the accelerating era at the redshift $z=0.83$ which shows a good compatibility with various current studies $0.5<z_t<1$ \cite{zt1,zt2,zt3,zt4,zt5,zt6}. As discussed in the previous section, it has been found that the reconstructed results of $\omega_{tot} (z)$ are in good agreement with recent observations. We also found that the interaction term $Q$ remains positive throughout the evolution and thus the energy transfers from THDE to DM, which is well consistent with the Le Chatelier-Braun principle and the second law of thermodynamics \cite{lcbpavon}. Furthermore, we shown the evolution of the normalized Hubble parameter $h(z)$ for the THDE model and compared that with the latest cosmic chronometer data.  We found that the above model reproduces the observed values of $h(z)$ quite well for each data point.\\

\par We also found that the squared sound speed $v^{2}<0$ which implies that the THDE model with Hubble cutoff is unstable against perturbation. In this context, it is worth mentioning that the cosmological constant case ($\omega_{\Lambda}=-1$) only presents a pure adiabatic behavior. On the other hand, the DE-DM interaction scheme leads to non-adiabatic cosmic evolution. In Ref. \cite{dedmnonad}, the authors shown that the constant and dynamical EoS parameter ($\omega \neq -1$) can give a different behavior at the structure formation level if treated as non-adiabatic or adiabatic components. However, they have also shown that the non-adiabatic DE models tend to overlap with the standard $\Lambda$CDM scenario at first order in linear perturbations. However, we close this work by mentioning that there are additional investigations require before the present model can be considered as a successful candidate for the description of DE. Therefore, we conclude that it would be interesting to examine the present scenario by considering other IR cutoffs. Further, it would also be interesting to perform non-adiabatic DE perturbation in order to reveal the non-adiabatic features in the THDE sector. This may alter the properties of THDE. We leave this task for future investigation. 
 \bibliographystyle{unsrtant}

\end{document}